\documentclass[showpacs]{revtex4}
\usepackage{indentfirst}
\begin{document}
\title{\bf Heavy hypernuclei with $A=3$ in a relativistic quark-gluon model}
\author{S.M. Gerasyuta}
\email{gerasyuta@SG6488.spb.edu}
\author{E.E. Matskevich}
\email{matskev@pobox.spbu.ru}
\affiliation{Department of Physics, St. Petersburg State Forest Technical
University, Institutski Per. 5, St. Petersburg 194021, Russia}
\begin{abstract}
We generalized our approach to the hypernuclei with $A=B=3$
containing one charm or one bottom quark.
We derive the relativistic nine-quark equations using the
dispersion relation technique. The hypernuclei as the system of
interacting quarks and gluons are considered. The relativistic nine-quark
amplitudes of hypernuclei, including the constituent quarks with the
charm or bottom are calculated. The approximate solutions of these
equations are obtained using a method based on the extraction
of leading singularities of the amplitudes. The poles of the
multiquark amplitudes allow us to determine the masses and the
binding energy of hypernuclei with the $A=3$. We predict the mass
spectrum of hypernuclei with $A=3$, which is valuable to further
experimental study of the hypernuclei with charm and bottom.
\end{abstract}
\maketitle
\section{Introduction.}
Due to the increasing computational resources becoming available to the field,
progress is being made toward a direct connection between QCD and nuclear physics
using the numerical technique of lattice QCD (LQCD). Steady progress is being
made toward this objective, but calculations at the physical light-quark masses
have not yet been performed, and essentially only one lattice spacing
has been used in calculation.

One reason that there are presently few LQCD calculations of $NN$ interacting
is the significantly greater complexity of multi-nucleon systems as compared
with systems of single mesons and baryons. A second reason is that significant
computational resources are required to generate high-quality ensembles of
gauge field configurations at or near the physical light-quark masses,
an effort that has only become practical with the availability of petascale
computers, and as yet these ensembles are not at sufficiently large volume
to be of use in nuclear physics \cite{1, 2, 3, 4, 5, 6, 7, 8, 9, 10, 11, 12, 13, 14}.

In our paper \cite{15} the $ ^3 He$ also as the system of interacting quarks
and gluons is considered. The relativistic nine-quark equations are found
in the framework of the dispersion relation technique.
The relativistic nine-quark amplitudes of $ ^3 He$, including the
$u$, $d$ quarks are calculated. The approximate solutions
of these equations using the method based on the extraction of leading
singularities of the amplitudes were obtained. The pole of the nonaquark
amplitudes determined the mass of $ ^3 He$.

Hypernuclei spectroscopy is enjoying an experimental renaissance with
ongoing and planned program at DA$\Phi$NE, FAIR, Jefferson Lab, J-PARC
and Mainz providing motivation for enhanced theoretical efforts
(for a recent review, see Ref. \cite{16}).

There are a number of theoretical predictions using various models:
the chiral $SU(3)$ quark model \cite{17}, the flavor $SU(3)$ skyrmion
model \cite{18}, the quark-delocation model \cite{19, 20} and the quark
cluster model \cite{21, 22}.

Lomon predicted a deuteron-like dibaryon resonance using R-matrix theory
\cite{23}. During the last several years, there has been a substantial
effort to determine the $NN$ interactions directly from QCD using the numeral
technique of LQCD. However, there is still little understanding of the
connection to the underlying theory of the strong interactions, QCD, and
the basic building blocks of nature, quarks and gluons. The main element in
connecting QCD to nuclei is determined by the properties of hadrons, and
can be described and predicted in terms of quark and gluon distribution.
The calculated scattering lengths and effective
ranges indicate that the pion is not the dominant contribution to the long
range part of the nuclear force at these large light-quark masses, as
anticipated from the single-hadron spectrum. This suggests that the
form of the nuclear interactions, and the effective potentials that
will reproduce the scattering amplitude below the inelastic threshold,
is qualitatively similar to be phenomenological potentials describing
the experimental scattering data at the physical pion mass.

In the recent paper \cite{24} the relativistic
six-quark equations are found in the framework of coupled-channel formalism.
The dynamical mixing between the subamplitudes of hexaquark is considered.
The six-quark amplitudes of dibaryons are calculated. The poles of these
amplitudes determine the masses of dibaryons. We calculated the contributions
of six-quark amplitudes to the hexaquark amplitudes.

In the previous paper \cite{25} the lowest hypernuclei are
considered in the framework of the dispersion relation technique.
The approximate solutions of the nine-quark equations using the method
based on the extraction of leading singularities of the amplitude are
obtained. The relativistic nine-quark amplitudes of low-lying hypernuclei,
including the three flavors ($u$, $d$, $s$) are calculated. The poles
of these amplitudes determine the masses of the hypernuclei with the
atomic (baryon) number $A=B=3$.

In Sec. II the relativistic nine-quark equations are derived for the
hypernuclei with $A=B=3$ containing one charm or one bottom quark.
The dynamical mixing between the subamplitudes of
hypernuclei is considered. Sec. III is devoted to the calculation
results for the masses of the low-lying charmed and bottom hypernuclei
(Table \ref{tab1}, \ref{tab2}).

In conclusion, the status of this model is discussed.

\section{Nine-quark amplitudes of hypernuclei.}
In the paper \cite{25} the lowest hypernuclei with the atomic (baryon)
number $A=B=3$: $ ^3_{Y}H$, $ ^3_{Y}He$, $nnY$,
where $Y=\Lambda$, $\Sigma_0$, $\Sigma_+$, $\Sigma_-$ are
considered in the dispersion relation technique.
The approximate solutions of the nine-quark equations using the method
based on the extraction of leading singularities of the amplitude are
obtained. The relativistic nonaquark amplitudes of low-lying hypernuclei,
including the three flavors ($u$, $d$, $s$) are calculated. The poles
of these amplitudes determine the masses of the hypernuclei.

In the present paper we generalize our method to the hypernuclei
with $A=B=3$ containing one charm or one bottom quark.
The relativistic nine-quark equations are derived using the
dispersion relation technique. These are the relativistic generalization
of the Faddeev-Yakubovsky \cite{26, 27} approach.

In constructing the equations for these nine-quark states the two-particle
interactions of quarks are considered. We use the results of the
paper \cite{28} in which the amplitudes of quark-quark interactions
$qq\rightarrow qq$ $a_n(s_{ik})$ are calculated in the framework of the
dispersion $N/D$ method with the input four-fermion interaction with
quantum numbers of the gluon by using iteration bootstrap procedure:

\begin{equation}
\label{1}
a_n(s_{ik})=\frac{G^2_n(s_{ik})}
{1-B_n(s_{ik})} \, \end{equation}

\noindent
where, $s_{ik}$ is the two-particle subenergy squared, $n$ corresponds to the quantum numbers of channel,
$G_n(s_{ik})$ are the quark-quark vertex functions (Table \ref{tab3}), and $B_n(s_{ik})$
is the Chew-Mandelstam function \cite{29}:

\begin{eqnarray}
\label{2}
B_n(s_{ik})=\int\limits_{(m_i+m_k)^2}^{\Lambda_n}
\, \frac{ds'_{ik}}{\pi}\frac{\rho_n(s'_{ik})G^2_n(s'_{ik})}
{s'_{ik}-s_{ik}} \, .
\end{eqnarray}

\noindent
$\rho_n (s_{ik})$ is the phase space:

\begin{eqnarray}
\label{3}
\rho_n (s_{ik})&=&\left(\alpha_n \frac{s_{ik}}{(m_i+m_k)^2}
+\beta_n+\delta_n \frac{(m_i-m_k)^2}{s_{ik}}\right)
\nonumber\\
&&\nonumber\\
&\times & \frac{\sqrt{(s_{ik}-(m_i+m_k)^2)(s_{ik}-(m_i-m_k)^2)}}
{s_{ik}} \, .
\end{eqnarray}

The coefficients $\alpha_n$, $\beta_n$ and $\delta_n$ are given in Table \ref{tab3}.
$n=1$ corresponds to $qq$-pairs with $J^P=0^+$, $n=2$ describes
the $qq$ pairs with $J^P=1^+$.

The current generates a nine-quark system. The correct equations for the
amplitude are obtained by taking into account all possible subamplitudes.
Then one should represent a nine-particle amplitude as a sum of the
following subamplitudes:

\begin{eqnarray}
\label{4}
A=\sum\limits_i A_1^i+\sum\limits_{i,j} A_2^{ij}+\sum\limits_{i,j,k} A_3^{ijk}\, .
\end{eqnarray}

In this paper, the total amplitude of nine-quark states is represented as the sum of
subamplitudes $A_1^i$, $A_2^{ij}$ and $A_3^{ijk}$, where $i$, $j$, $k$ are quantum
numbers of diquarks. $A_1^i$ is amplitude of diquark
and seven quarks. $A_2^{ij}$ corresponds two diquarks
and five quarks. Finally, $A_3^{ijk}$ is just the
three baryon state.

We consider only the planar diagrams, the other diagrams are neglected due to the rules
of $1/N_c$ expansion \cite{30, 31, 32}.
The total amplitude can be represented graphically as a sum of diagrams.
For example, we use Fig. 1 with the graphical equation for the
reduced amplitude $\alpha_3^{1^{uu}1^{uu}0^{dc}}$ for the state $ ^3_{\Sigma_c^0} He$ $pp\Sigma_c^0$.

First, we have to draw all graphical equations for all amplitudes taking into account
all possible rescattering, then the analytical expressions are constructed.

The system of integral equations is solved by approximate methods.
We extract the two-particle singularities and the 3th, 4th, 5th, 6th partial
singularities take into account in the coupled equations.
In the considered example
(amplitude  $\alpha_3^{1^{uu}1^{uu}0^{dc}}$ for the state $ ^3_{\Sigma_c^0} He$ $pp\Sigma_c^0$
with the isospin projection $I_3=0$ and the spin-parity $J^P=\frac{1}{2}^+$) the reduced
amplitudes are constructed:

\begin{eqnarray}
\label{5}
\alpha_3^{1^{uu}1^{uu}0^{dc}}&=&\lambda+4\, \alpha_1^{1^{uu}} I_9(1^{uu}1^{uu}0^{dc}1^{uu})
+3\, \alpha_1^{1^{dd}} I_{12}(0^{dc}1^{uu}1^{uu}1^{dd})+\alpha_1^{0^{ud}} (4\, I_9(1^{uu}0^{dc}1^{uu}0^{ud})
+12\, I_{12}(1^{uu}1^{uu}0^{dc}0^{ud}))\nonumber\\
&+&4\, \alpha_1^{0^{uc}} I_9(1^{uu}0^{dc}1^{uu}0^{uc})
+3\, \alpha_1^{0^{dc}} I_{12}(0^{dc}1^{uu}1^{uu}0^{dc})+24\, \alpha_2^{1^{uu}0^{ud}} I_{15}(1^{uu}1^{uu}0^{dc}1^{uu}0^{ud})\nonumber\\
&+&24\, \alpha_2^{1^{dd}0^{ud}} I_{14}(1^{uu}0^{dc}1^{uu}0^{ud}1^{dd})
+12\, \alpha_2^{1^{dd}0^{uc}} I_{15}(1^{uu}0^{dc}1^{uu}0^{uc}1^{dd})
+6\, \alpha_2^{1^{dd}0^{dc}} I_{13}(0^{dc}1^{uu}1^{uu}1^{dd}0^{dc})\nonumber\\
&+&\alpha_2^{0^{ud}0^{ud}} (12\, I_{13}(1^{uu}1^{uu}0^{dc}0^{ud}0^{ud})+24\, I_{14}(1^{uu}1^{uu}0^{dc}0^{ud}0^{ud})
+12\, I_{15}(0^{dc}1^{uu}1^{uu}0^{ud}0^{ud}))\nonumber\\
&+&12\, \alpha_2^{0^{ud}0^{uc}} I_{15}(0^{dc}1^{uu}1^{uu}0^{uc}0^{ud})
+\alpha_2^{0^{ud}0^{dc}} (24\, I_{14}(1^{uu}0^{dc}1^{uu}0^{ud}0^{dc})+12\, I_{15}(1^{uu}0^{dc}1^{uu}0^{ud}0^{dc}))\nonumber\\
&+&12\, \alpha_3^{1^{dd}0^{ud}0^{ud}} I_{18}(0^{dc}1^{uu}1^{uu}1^{dd}0^{ud}0^{ud})
+\alpha_3^{0^{ud}0^{ud}0^{dc}} (24\, I_{16}(1^{uu}0^{dc}1^{uu}0^{ud}0^{ud}0^{dc})\nonumber\\
&+&12\, I_{18}(0^{dc}1^{uu}1^{uu}0^{dc}0^{ud}0^{ud}))
\end{eqnarray}

We consider the system of 23 equations in the case of $ ^3_{\Sigma_c^0} He$ ($pp\Sigma_c^0$) state:

\begin{eqnarray}
\label{6}
\alpha_1: & & \, \, 1^{uu}, \, 1^{dd}, \, 0^{ud}, \, 0^{uc}, \, 0^{dc}
\\
\label{7}
\alpha_2: & & \, \, 1^{uu}1^{uu}, \, 1^{uu}1^{dd}, \, 1^{uu}0^{ud},
 \, 1^{uu}0^{uc}, \, 1^{uu}0^{dc}, \nonumber
\\
 & & \, \, 1^{dd}1^{dd}, \, 1^{dd}0^{ud}, \, 1^{dd}0^{uc}, \, 1^{dd}0^{dc}, \nonumber
\\
 & & \, \, 0^{ud}0^{ud}, \, 0^{ud}0^{uc}, \, 0^{ud}0^{dc}
 \\
\label{8}
\alpha_3: & & \, \, 1^{uu}1^{uu}1^{dd}, \, 1^{uu}1^{uu}0^{dc},
 \nonumber
\\
 & & \, \, 1^{uu}0^{ud}1^{dd}, \, 1^{uu}0^{ud}0^{dc}, \nonumber
\\
 & & \, \, 0^{ud}0^{ud}1^{dd}, \, 0^{ud}0^{ud}0^{dc}
\end{eqnarray}

The coefficients of the coupled equations are determined by the
permutation of quarks (Appendix A). The functions $I_1$ -- $I_{18}$ ((\ref{B1}) -- (\ref{B16}) in the Appendix B)
take into account the interactions of quarks and gluons.

The main contributions are calculated using the functions $I_1$ and $I_2$:

\begin{eqnarray}
\label{9}
I_1(ij)&=&\frac{B_j(s_0^{13})}{B_i(s_0^{12})}
\int\limits_{(m_1+m_2)^2}^{\Lambda\frac{(m_1+m_2)^2}{4}}
\frac{ds'_{12}}{\pi}\frac{G_i^2(s_0^{12})\rho_i(s'_{12})}
{s'_{12}-s_0^{12}} \int\limits_{-1}^{+1} \frac{dz_1(1)}{2}
\frac{1}{1-B_j (s'_{13})}\, , \\
&&\nonumber\\
\label{10}
I_2(ijk)&=&\frac{B_j(s_0^{13}) B_k(s_0^{24})}{B_i(s_0^{12})}
\int\limits_{(m_1+m_2)^2}^{\Lambda\frac{(m_1+m_2)^2}{4}}
\frac{ds'_{12}}{\pi}\frac{G_i^2(s_0^{12})\rho_i(s'_{12})}
{s'_{12}-s_0^{12}}
\frac{1}{2\pi}\int\limits_{-1}^{+1}\frac{dz_1(2)}{2}
\int\limits_{-1}^{+1} \frac{dz_2(2)}{2}\nonumber\\
&&\nonumber\\
&\times&
\int\limits_{z_3(2)^-}^{z_3(2)^+} dz_3(2)
\frac{1}{\sqrt{1-z_1^2(2)-z_2^2(2)-z_3^2(2)+2z_1(2) z_2(2) z_3(2)}}
\nonumber\\
&&\nonumber\\
&\times& \frac{1}{1-B_j (s'_{13})} \frac{1}{1-B_k (s'_{24})}
 \, ,
\end{eqnarray}

\noindent
where $i$, $j$, $k$ correspond to the diquarks.

\newpage

\begin{table}
\caption{Masses of charmed hypernuclei. Parameters of model: $\Lambda=9.0$,
$g=0.2122$, $m_{u, d}=495\, MeV$, $m_c=1655\, MeV$.}\label{tab1}
\begin{tabular}{|l|c|c|c|c|c|c|c|}
\hline
hypernuclei & quark content & $Q$ & $I_3$ & $I$ & $J^P$ & mass, MeV & binding energy, MeV \\ [5pt]
\hline
$pn\Lambda_c$ $( ^3_{\Lambda_c} H)$          & $uud \, udd \, udc$ & +2 & 0     & 0, 1    & $\frac{1}{2}^+$, $\frac{3}{2}^+$ & 4102 & 63  \\ [5pt]
$pn\Sigma_c^+$ $( ^3_{\Sigma_c^+} H)$        & $uud \, udd \, udc$ & +2 & 0     & 0, 1, 2 & $\frac{1}{2}^+$, $\frac{3}{2}^+$ & 4102 & 229 \\ [5pt]
$pn\Sigma_c^{++}$ $( ^3_{\Sigma_c^{++}} H)$  & $uud \, udd \, uuc$ & +3 & +1    & 1, 2    & $\frac{1}{2}^+$, $\frac{3}{2}^+$ & 4105 & 227 \\ [5pt]
$pn\Sigma_c^0$ $( ^3_{\Sigma_c^0} H)$        & $uud \, udd \, ddc$ & +1 & $-1$  & 1, 2    & $\frac{1}{2}^+$, $\frac{3}{2}^+$ & 4105 & 227 \\ [5pt]
$pp\Lambda_c$ $( ^3_{\Lambda_c} He)$         & $uud \, uud \, udc$ & +3 & +1    & 1       & $\frac{1}{2}^+$                  & 4104 & 59  \\ [5pt]
$nn\Lambda_c$                                & $udd \, udd \, udc$ & +1 & $-1$  & 1       & $\frac{1}{2}^+$                  & 4104 & 63  \\ [5pt]
$pp\Sigma_c^+$ $( ^3_{\Sigma_c^+} He)$       & $uud \, uud \, udc$ & +3 & +1    & 1, 2    & $\frac{1}{2}^+$                  & 4104 & 225 \\ [5pt]
$nn\Sigma_c^+$                               & $udd \, udd \, udc$ & +1 & $-1$  & 1, 2    & $\frac{1}{2}^+$                  & 4104 & 229 \\ [5pt]
$pp\Sigma_c^{++}$ $( ^3_{\Sigma_c^{++}} He)$ & $uud \, uud \, uuc$ & +4 & +2    & 2       & $\frac{1}{2}^+$                  & 4044 & 286 \\ [5pt]
$nn\Sigma_c^0$                               & $udd \, udd \, ddc$ & 0  & $-2$  & 2       & $\frac{1}{2}^+$                  & 4044 & 290 \\ [5pt]
$pp\Sigma_c^0$ $( ^3_{\Sigma_c^0} He)$       & $uud \, uud \, ddc$ & +2 & 0     & 0, 1, 2 & $\frac{1}{2}^+$                  & 4144 & 186 \\ [5pt]
$nn\Sigma_c^{++}$                            & $udd \, udd \, uuc$ & +2 & 0     & 0, 1, 2 & $\frac{1}{2}^+$                  & 4144 & 190 \\ [5pt]
\hline
\end{tabular}
\end{table}

\begin{table}
\caption{Masses of bottom hypernuclei. Parameters of model: $\Lambda=9.0$, $\Lambda_b=4.43$,
$g=0.1682$, $m_{u, d}=495\, MeV$, $m_b=4840\, MeV$.}\label{tab2}
\begin{tabular}{|l|c|c|c|c|c|c|c|}
\hline
hypernuclei & quark content & $Q$ & $I_3$ & $I$ & $J^P$ & mass, MeV & binding energy, MeV \\ [5pt]
\hline
$pn\Lambda_b$ $( ^3_{\Lambda_b} H)$    & $uud \, udd \, udb$ & +1   & 0     & 0, 1    & $\frac{1}{2}^+$, $\frac{3}{2}^+$ & 7330 & 168  \\ [5pt]
$pn\Sigma_b^0$ $( ^3_{\Sigma_b^0} H)$  & $uud \, udd \, udb$ & +1   & 0     & 0, 1, 2 & $\frac{1}{2}^+$, $\frac{3}{2}^+$ & 7330 &  \\ [5pt]
$pn\Sigma_b^+$ $( ^3_{\Sigma_b^+} H)$  & $uud \, udd \, uub$ & +2   & +1    & 1, 2    & $\frac{1}{2}^+$, $\frac{3}{2}^+$ & 7366 & 323 \\ [5pt]
$pn\Sigma_b^-$ $( ^3_{\Sigma_b^-} H)$  & $uud \, udd \, ddb$ & 0    & $-1$  & 1, 2    & $\frac{1}{2}^+$, $\frac{3}{2}^+$ & 7366 & 328 \\ [5pt]
$pp\Lambda_b$ $( ^3_{\Lambda_b} He)$   & $uud \, uud \, udb$ & +2   & +1    & 1       & $\frac{1}{2}^+$                  & 7366 & 130  \\ [5pt]
$nn\Lambda_b$                          & $udd \, udd \, udb$ & 0    & $-1$  & 1       & $\frac{1}{2}^+$                  & 7366 & 134  \\ [5pt]
$pp\Sigma_b^0$ $( ^3_{\Sigma_b^0} He)$ & $uud \, uud \, udb$ & +2   & +1    & 1, 2    & $\frac{1}{2}^+$                  & 7366 &  \\ [5pt]
$nn\Sigma_b^0$                         & $udd \, udd \, udb$ & 0    & $-1$  & 1, 2    & $\frac{1}{2}^+$                  & 7366 &  \\ [5pt]
$pp\Sigma_b^+$ $( ^3_{\Sigma_b^+} He)$ & $uud \, uud \, uub$ & +3   & +2    & 2       & $\frac{1}{2}^+$                  & 7364 & 323 \\ [5pt]
$nn\Sigma_b^-$                         & $udd \, udd \, ddb$ & $-1$ & $-2$  & 2       & $\frac{1}{2}^+$                  & 7364 & 332 \\ [5pt]
$pp\Sigma_b^-$ $( ^3_{\Sigma_b^-} He)$ & $uud \, uud \, ddb$ & +1   & 0     & 0, 1, 2 & $\frac{1}{2}^+$                  & 7400 & 292 \\ [5pt]
$nn\Sigma_b^+$                         & $udd \, udd \, uub$ & +1   & 0     & 0, 1, 2 & $\frac{1}{2}^+$                  & 7400 & 291 \\ [5pt]
\hline
\end{tabular}
\end{table}

\begin{table}
\caption{The vertex functions and coefficients of Chew-Mandelstam functions.}\label{tab3}
\begin{tabular}{|c|c|c|c|c|c|}
\hline
\, $n$ \, & \, $J^P$ \, & $G_n^2(s_{kl})$ & \, $\alpha_n$ \, & $\beta_n$ & \, $\delta_n$ \, \\
\hline
& & & & & \\
1 & $0^+$ & $\frac{4g}{3}-\frac{8g m_{kl}^2}{(3s_{kl})}$
& $\frac{1}{2}$ & $-\frac{1}{2}\frac{(m_k-m_l)^2}{(m_k+m_l)^2}$ & $0$ \\
& & & & & \\
2 & $1^+$ & $\frac{2g}{3}$ & $\frac{1}{3}$
& $\frac{4m_k m_l}{3(m_k+m_l)^2}-\frac{1}{6}$
& $-\frac{1}{6}$ \\
& & & & & \\
\hline
\end{tabular}
\end{table}

\section{Calculation results.}
The poles of the reduced amplitudes $\alpha_1$, $\alpha_2$, $\alpha_3$
correspond to the bound states and determine the masses of
hypernuclei with the atomic number $A=3$ (Tables \ref{tab1}, \ref{tab2}).

The values of quark masses in the our calculations are the usual for the
heavy baryons in our model: $m_{u, d}=495\, MeV$, $m_c=1655\, MeV$ and $m_b=4840\, MeV$.
For the charmed hypernuclei we use the two parameters: the gluon coupling constant $g=0.2122$,
and the cutoff $\Lambda=9$ which are similar to the case of light hypernuclei.

The experimental data for the bottom hypernuclei are absent, therefore
the mass of the heaviest bottom state $ ^3_{\Sigma_b^-} He$
($pp\Sigma_b^-$) is described to be $7400\, MeV$.
This differs from threshold value for this state by more than $200\, MeV$.
This allows us to determine the value of gluon coupling constant $g=0.1682$
for the case of bottom hypernuclei. Another two parameters $\Lambda=9$ and $\Lambda_b=4.43$
are taken similar to the ones of previous papers \cite{25, 33}.
The calculation results are represented in the Tables \ref{tab1}, \ref{tab2}.

\section{Conclusions.}
In the papers \cite{24, 34} we have developed the method of studying of
hadrons and hypernuclei. This is the dispersion relations approach using
the main postulates of $S$-matrix method. The relativistic generalizations
of the three-body Faddeev equations in the form of dispersion relations
is considered. The mass spectrum of the $S$-wave baryons, including $u$, $d$, $s$ quarks
was calculated by a method based on isolating the leading singularities in the amplitude.
The approximate solution of the integral three-quark equations by taking into
account the two-particle and triangle singularities is calculated.
If we consider such approximation and define all
the smooth functions of the subenergy variables (as compared with the
singular part of the amplitude) in the middle point of the physical region
of Dalitz-plot, the problem is reduced to the one of solving the system
of algebraic equations.

In the present paper, the relativistic nine-quark equations in the framework of
dispersion relation technique are derived. We predict the new hypernuclei with
the one $c$- or $b$-quark. The states are calculated using the isospin $I=0,1,2$
and the spin-parity $J^P=\frac{1}{2}^+$, $\frac{3}{2}^+$. The masses of nine-quark
hypernuclei with $A=3$ are degenerated. The electromagnetic effect is not included.

\begin{acknowledgments}
The reported study was partially supported by RFBR, research project No. 13-02-91154.
\end{acknowledgments}

\newpage

\appendix

\section{}

In Fig. 1 the coefficient of the term $I_{9}(1^{uu}1^{uu}0^{dc}1^{uu})\, \alpha_1^{1^{uu}}$
is equal to 4, that is the number $4=2$ (the permutation of particles 1 and 2) $\times 2$
(the permutation of particles 3 and 4); the coefficient of the term
$I_{12}(0^{dc}1^{uu}1^{uu}1^{dd})\, \alpha_1^{1^{dd}}$ is equal to 3
(we can replace the 7-th $d$-quark with the 8-th and 9-th $d$-quarks);
the coefficient of the term $I_{13}(0^{dc}1^{uu}1^{uu}1^{dd}0^{dc})\, \alpha_2^{1^{dd}0^{dc}}$
is equal to 6, that is the number $6=3\times 2$ (we can replace the 7-th $d$-quark with the 8-th
and 9-th $d$-quarks, and then replace the 8-th $d$-quark with the 9-th quark);
the coefficient of the term $I_{14}(1^{uu}1^{uu}0^{dc}0^{ud}0^{ud})\, \alpha_2^{0^{ud}0^{ud}}$
is equal to 24, that is the number $24=2$ (the permutation of particles 1 and 2) $\times 2$
(the permutation of particles 3 and 4) $\times 3\times 2$ (we can replace the 7-th $d$-quark with
the 8-th and 9-th $d$-quarks, and then replace the 8-th $d$-quark with the 9-th quark);
the coefficient of the term $I_{15}(0^{dc}1^{uu}1^{uu}0^{uc}0^{ud})\, \alpha_2^{0^{ud}0^{uc}}$
is equal to 12, that is the number $12=2$ (the permutation of particles 3 and 4) $\times 3$
(we can replace the 7-th $d$-quark with the 8-th and 9-th $d$-quarks) $\times 2$
(the permutation of pairs (34) and (56)); the coefficient of the term
$I_{16}(1^{uu}0^{dc}1^{uu}0^{ud}0^{ud}0^{dc})\, \alpha_3^{0^{ud}0^{ud}0^{dc}}$
is equal to 24, that is the number $24=2$ (the permutation of particles 1 and 2)
$\times 3\times 2$ (we can replace the 7-th $d$-quark with the 8-th and 9-th $d$-quarks,
and then replace the 8-th $d$-quark with the 9-th quark) $\times 2$
(the permutation of pairs (12) and (56)); the coefficient of the term
$I_{18}(0^{dc}1^{uu}1^{uu}0^{dc}0^{ud}0^{ud})\, \alpha_3^{0^{ud}0^{ud}0^{dc}}$
is equal to 12, that is the number $12=2$ (the permutation of pairs (34) and (56))
$\times 3\times 2$ (we can replace the 7-th $d$-quark with the 8-th and
9-th $d$-quarks, and then replace the 8-th $d$-quark with the 9-th quark).

The similar approach allows us to take into account the coefficients
in all the diagrams and equations.


\section{}

We used the functions
$I_1$, $I_2$, $I_3$, $I_4$, $I_5$, $I_6$, $I_7$, $I_8$, $I_9$, $I_{11}$, $I_{12}$, $I_{13}$,
$I_{14}$, $I_{15}$, $I_{16}$, $I_{18}$:

\begin{eqnarray}
\label{B1}
I_1(ij)&=&\frac{B_j(s_0^{13})}{B_i(s_0^{12})}
\int\limits_{(m_1+m_2)^2}^{\frac{(m_1+m_2)^2\Lambda_i}{4}}
\frac{ds'_{12}}{\pi}\frac{G_i^2(s_0^{12})\rho_i(s'_{12})}
{s'_{12}-s_0^{12}} \int\limits_{-1}^{+1} \frac{dz_1(1)}{2}
\frac{1}{1-B_j (s'_{13})}\, ,
\end{eqnarray}

\begin{eqnarray}
\label{B2}
I_2(ijk)&=&\frac{B_j(s_0^{13}) B_k(s_0^{24})}{B_i(s_0^{12})}
\int\limits_{(m_1+m_2)^2}^{\frac{(m_1+m_2)^2\Lambda_i}{4}}
\frac{ds'_{12}}{\pi}\frac{G_i^2(s_0^{12})\rho_i(s'_{12})}
{s'_{12}-s_0^{12}}
\frac{1}{2\pi}\int\limits_{-1}^{+1}\frac{dz_1(2)}{2}
\int\limits_{-1}^{+1} \frac{dz_2(2)}{2}\nonumber\\
&&\nonumber\\
&\times&
\int\limits_{z_3(2)^-}^{z_3(2)^+} dz_3(2)
\frac{1}{\sqrt{1-z_1^2(2)-z_2^2(2)-z_3^2(2)+2z_1(2) z_2(2) z_3(2)}}
\nonumber\\
&&\nonumber\\
&\times& \frac{1}{1-B_j (s'_{13})} \frac{1}{1-B_k (s'_{24})}
 \, ,
\end{eqnarray}

\begin{eqnarray}
\label{B3}
I_3(ijk)&=&\frac{B_k(s_0^{23})}{B_i(s_0^{12}) B_j(s_0^{34})}
\int\limits_{(m_1+m_2)^2}^{\frac{(m_1+m_2)^2\Lambda_i}{4}}
\frac{ds'_{12}}{\pi}\frac{G_i^2(s_0^{12})\rho_i(s'_{12})}
{s'_{12}-s_0^{12}}\nonumber\\
&&\nonumber\\
&\times&\int\limits_{(m_3+m_4)^2}^{\frac{(m_3+m_4)^2\Lambda_j}{4}}
\frac{ds'_{34}}{\pi}\frac{G_j^2(s_0^{34})\rho_j(s'_{34})}
{s'_{34}-s_0^{34}}
\nonumber\\
&&\nonumber\\
&\times&
\int\limits_{-1}^{+1} \frac{dz_1(3)}{2} \int\limits_{-1}^{+1}
\frac{dz_2(3)}{2} \frac{1}{1-B_k (s'_{23})} \, ,
\end{eqnarray}

\begin{eqnarray}
\label{B4}
I_4(ijk)&=&I_1(ik) \, ,
\end{eqnarray}

\begin{eqnarray}
\label{B5}
I_5(ijkl)&=&I_2(ikl) \, ,
\end{eqnarray}

\begin{eqnarray}
\label{B6}
I_6(ijkl)&=&I_1(ik) \times I_1(jl) \, ,
\end{eqnarray}

\begin{eqnarray}
\label{B7}
I_7(ijkl)&=&\frac{B_k(s_0^{23})B_l(s_0^{45})}{B_i(s_0^{12}) B_j(s_0^{34})}
\int\limits_{(m_1+m_2)^2}^{\frac{(m_1+m_2)^2\Lambda_i}{4}}
\frac{ds'_{12}}{\pi}\frac{G_i^2(s_0^{12})\rho_i(s'_{12})}
{s'_{12}-s_0^{12}}\nonumber\\
&&\nonumber\\
&\times&\int\limits_{(m_3+m_4)^2}^{\frac{(m_3+m_4)^2\Lambda_j}{4}}
\frac{ds'_{34}}{\pi}\frac{G_j^2(s_0^{34})\rho_j(s'_{34})}
{s'_{34}-s_{34}}
\frac{1}{2\pi}\int\limits_{-1}^{+1}\frac{dz_1(7)}{2}
\int\limits_{-1}^{+1} \frac{dz_2(7)}{2}
\int\limits_{-1}^{+1} \frac{dz_3(7)}{2}
\nonumber\\
&&\nonumber\\
&\times&
\int\limits_{z_4(7)^-}^{z_4(7)^+} dz_4(7)
\frac{1}{\sqrt{1-z_1^2(7)-z_3^2(7)-z_4^2(7)+2z_1(7) z_3(7) z_4(7)}}
\nonumber\\
&&\nonumber\\
&\times& \frac{1}{1-B_k (s'_{23})} \frac{1}{1-B_l (s'_{45})}\, ,
\end{eqnarray}

\begin{eqnarray}
\label{B8}
I_8(ijklm)&=&\frac{B_k(s_0^{15})B_l(s_0^{23})B_m(s_0^{46})}
{B_i(s_0^{12}) B_j(s_0^{34})}
\int\limits_{(m_1+m_2)^2}^{\frac{(m_1+m_2)^2\Lambda_i}{4}}
\frac{ds'_{12}}{\pi}\frac{G_i^2(s_0^{12})\rho_i(s'_{12})}
{s'_{12}-s_0^{12}}\nonumber\\
&&\nonumber\\
&\times&\int\limits_{(m_3+m_4)^2}^{\frac{(m_3+m_4)^2\Lambda_j}{4}}
\frac{ds'_{34}}{\pi}\frac{G_j^2(s_0^{34})\rho_j(s'_{34})}
{s'_{34}-s_0^{34}}\nonumber\\
&&\nonumber\\
&\times&\frac{1}{(2\pi)^2}\int\limits_{-1}^{+1}\frac{dz_1(8)}{2}
\int\limits_{-1}^{+1} \frac{dz_2(8)}{2}
\int\limits_{-1}^{+1} \frac{dz_3(8)}{2}
\int\limits_{z_4(8)^-}^{z_4(8)^+} dz_4(8)
\int\limits_{-1}^{+1} \frac{dz_5(8)}{2}
\int\limits_{z_6(8)^-}^{z_6(8)^+} dz_6(8)
\nonumber\\
&&\nonumber\\
&\times&
\frac{1}{\sqrt{1-z_1^2(8)-z_3^2(8)-z_4^2(8)+2z_1(8) z_3(8) z_4(8)}}
\nonumber\\
&&\nonumber\\
&\times&
\frac{1}{\sqrt{1-z_2^2(8)-z_5^2(8)-z_6^2(8)+2z_2(8) z_5(8) z_6(8)}}
\nonumber\\
&&\nonumber\\
&\times& \frac{1}{1-B_k (s'_{15})} \frac{1}{1-B_l (s'_{23})}
\frac{1}{1-B_m (s'_{46})} \, ,
\end{eqnarray}

\begin{eqnarray}
\label{B9}
I_9(ijkl)&=&I_3(ijl) \, ,
\end{eqnarray}

\begin{eqnarray}
\label{B10}
I_{11}(ijklm)&=&I_1(ik) \times I_2(jlm) \, ,
\end{eqnarray}

\begin{eqnarray}
\label{B11}
I_{12}(ijkl)&=&I_1(il) \, ,
\end{eqnarray}

\begin{eqnarray}
\label{B12}
I_{13}(ijklm)&=&I_2(ilm) \, ,
\end{eqnarray}

\begin{eqnarray}
\label{B13}
I_{14}(ijklm)&=&I_1(il) \times I_1(jm) \, ,
\end{eqnarray}

\begin{eqnarray}
\label{B14}
I_{15}(ijklm)&=&I_7(ijlm) \, ,
\end{eqnarray}

\begin{eqnarray}
\label{B15}
I_{16}(ijklmn)&=&I_8(ijlmn) \, ,
\end{eqnarray}

\begin{eqnarray}
\label{B16}
I_{18}(ijklmn)&=&I_1(il) \times I_2(jmn) \, .
\end{eqnarray}

\noindent
Here $i$, $j$, $k$, $l$, $m$, $n$, $p$, $q$ correspond to the diquarks with the
spin-parity $J^P=0^+, 1^+$.

The other functions $I_i$ can be neglected. The contributions of these functions
are smaller of few orders as compared the functions (\ref{B1}) -- (\ref{B16}). We do not take into account
these functions in the systems of coupled equations.

\newpage

\vskip60pt

\begin{picture}(600,90)
\put(-20,54){\line(1,0){18}}
\put(-20,53){\line(1,0){18}}
\put(-20,52){\line(1,0){18}}
\put(-20,51){\line(1,0){18}}
\put(-20,50){\line(1,0){18}}
\put(-20,49){\line(1,0){18}}
\put(-20,48){\line(1,0){18}}
\put(-20,47){\line(1,0){18}}
\put(-20,46){\line(1,0){18}}
\put(10,50){\circle{25}}
\put(-1,46){\line(1,1){15}}
\put(2,41){\line(1,1){17}}
\put(7.5,38.5){\line(1,1){14}}
\put(20,68){\circle{16}}
\put(31,50){\circle{16}}
\put(20,32){\circle{16}}
\put(6,61.5){\vector(1,3){10}}
\put(19,85){7}
\put(28,70){\vector(3,2){19}}
\put(50,80){1}
\put(28,70){\vector(3,-1){22}}
\put(53,63){2}
\put(39,50){\vector(3,2){14}}
\put(55,52){3}
\put(39,50){\vector(3,-2){14}}
\put(55,41){4}
\put(28,30){\vector(3,1){22}}
\put(53,30){5}
\put(28,30){\vector(3,-2){19}}
\put(50,13){6}
\put(6,39){\vector(1,-3){10}}
\put(19,10){8}
\put(1,42){\vector(1,-4){8.5}}
\put(0,9){9}
\put(14,65){\small $1^{uu}$}
\put(25,47){\small $1^{uu}$}
\put(14,29){\small $0^{dc}$}
\put(27,88){$d$}
\put(60,84){$u$}
\put(63,64){$u$}
\put(63,54){$u$}
\put(63,41){$u$}
\put(63,30){$d$}
\put(60,11){$c$}
\put(26,2){$d$}
\put(10,-2){$d$}
\put(74,48){=}
\put(-5,-25){$\alpha_3^{1^{uu}1^{uu}0^{dc}}$}
\put(87,54){\line(1,0){19}}
\put(87,53){\line(1,0){20}}
\put(87,52){\line(1,0){21}}
\put(87,51){\line(1,0){22}}
\put(87,50){\line(1,0){23}}
\put(87,49){\line(1,0){22}}
\put(87,48){\line(1,0){21}}
\put(87,47){\line(1,0){20}}
\put(87,46){\line(1,0){19}}
\put(121,67){\circle{16}}
\put(127,50){\circle{16}}
\put(121,32){\circle{16}}
\put(105,55){\vector(1,4){8.2}}
\put(117,86){7}
\put(129,70){\vector(1,1){13}}
\put(131,82){1}
\put(129,70){\vector(3,-1){18}}
\put(148,67){2}
\put(135,50){\vector(3,2){16}}
\put(155,55){3}
\put(135,50){\vector(3,-2){16}}
\put(155,39){4}
\put(129,30){\vector(3,1){18}}
\put(148,26){5}
\put(129,30){\vector(1,-1){13}}
\put(132,9){6}
\put(105,45){\vector(1,-4){8.2}}
\put(116,08){8}
\put(105,45){\vector(0,-1){34}}
\put(97,10){9}
\put(115,64){\small $1^{uu}$}
\put(121,47){\small $1^{uu}$}
\put(115,29){\small $0^{dc}$}
\put(125,92){$d$}
\put(140,89){$u$}
\put(158,70){$u$}
\put(164,60){$u$}
\put(164,37){$u$}
\put(158,24){$d$}
\put(145,8){$c$}
\put(122,-2){$d$}
\put(105,-2){$d$}
\put(178,48){+}
\put(115,-25){$\lambda$}
\put(197,54){\line(1,0){18}}
\put(197,53){\line(1,0){18}}
\put(197,52){\line(1,0){18}}
\put(197,51){\line(1,0){18}}
\put(197,50){\line(1,0){18}}
\put(197,49){\line(1,0){18}}
\put(197,48){\line(1,0){18}}
\put(197,47){\line(1,0){18}}
\put(197,46){\line(1,0){18}}
\put(227,50){\circle{25}}
\put(216,46){\line(1,1){15}}
\put(219,41){\line(1,1){17}}
\put(224.5,38.5){\line(1,1){14}}
\put(247,50){\circle{16}}
\put(280,60){\circle{16}}
\put(280,40){\circle{16}}
\put(227,30){\circle{16}}
\put(224,62){\vector(-1,4){7}}
\put(220,87){7}
\put(227,62){\vector(0,1){29}}
\put(229,87){8}
\put(230,62){\vector(1,4){7}}
\put(239,85){9}
\put(232,61){\vector(1,0){40}}
\put(255,66){1}
\put(255,50){\vector(3,2){17}}
\put(262,48){2}
\put(255,50){\vector(3,-2){17}}
\put(268,43){3}
\put(232,39){\vector(1,0){40}}
\put(255,28){4}
\put(288,61){\vector(3,1){20}}
\put(313,70){1}
\put(288,61){\vector(3,-1){20}}
\put(313,53){2}
\put(288,39){\vector(3,1){20}}
\put(313,41){3}
\put(288,39){\vector(3,-1){20}}
\put(313,24){4}
\put(227,22){\vector(2,-3){11}}
\put(240,10){5}
\put(227,22){\vector(-2,-3){11}}
\put(209,10){6}
\put(241,47){\small $1^{uu}$}
\put(274,57){\small $1^{uu}$}
\put(274,37){\small $1^{uu}$}
\put(221,27){\small $0^{dc}$}
\put(216,97){$d$}
\put(231,99){$d$}
\put(243,97){$d$}
\put(321,74){$u$}
\put(321,53){$u$}
\put(321,43){$u$}
\put(321,21){$u$}
\put(263,65){$u$}
\put(255,55){$u$}
\put(255,41){$u$}
\put(263,30){$u$}
\put(243,-1){$d$}
\put(207,-1){$c$}
\put(332,48){+}
\put(220,-25){$4\, I_{9}(1^{uu}1^{uu}0^{dc}1^{uu})\, \alpha_1^{1^{uu}}$}
\put(347,54){\line(1,0){18}}
\put(347,53){\line(1,0){18}}
\put(347,52){\line(1,0){18}}
\put(347,51){\line(1,0){18}}
\put(347,50){\line(1,0){18}}
\put(347,49){\line(1,0){18}}
\put(347,48){\line(1,0){18}}
\put(347,47){\line(1,0){18}}
\put(347,46){\line(1,0){18}}
\put(377,50){\circle{25}}
\put(366,46){\line(1,1){15}}
\put(369,41){\line(1,1){17}}
\put(374.5,38.5){\line(1,1){14}}
\put(390,66){\circle{16}}
\put(423,62){\circle{16}}
\put(394,39){\circle{16}}
\put(372,30.5){\circle{16}}
\put(371,60.5){\vector(1,4){9}}
\put(382,90){8}
\put(376,62){\vector(1,2){16}}
\put(394,85){9}
\put(396,71){\vector(1,1){16}}
\put(403,87){7}
\put(396,71){\vector(2,-1){18.5}}
\put(406,68){1}
\put(389,53){\vector(3,1){25.5}}
\put(403,50){2}
\put(431,62){\vector(2,1){20}}
\put(457,68){1}
\put(431,62){\vector(2,-1){20}}
\put(457,50){2}
\put(400.5,34){\vector(4,-1){21}}
\put(424,25){3}
\put(400.5,34){\vector(1,-4){5.2}}
\put(408,7){4}
\put(371,22.5){\vector(2,-3){12}}
\put(386,3){5}
\put(371,22.5){\vector(-2,-3){12}}
\put(353,3){6}
\put(384,63){\small $1^{dd}$}
\put(417,59){\small $0^{dc}$}
\put(388,36){\small $1^{uu}$}
\put(366,27.5){\small $1^{uu}$}
\put(381,103){$d$}
\put(394,100){$d$}
\put(414,93){$d$}
\put(465,73){$d$}
\put(465,47){$c$}
\put(412,73){$d$}
\put(411,49){$c$}
\put(432,20){$u$}
\put(402,-1){$u$}
\put(386,-8){$u$}
\put(352,-8){$u$}
\put(357,-25){$3\, I_{12}(0^{dc}1^{uu}1^{uu}1^{dd})\, \alpha_1^{1^{dd}}$}
\end{picture}

\vskip110pt

\begin{picture}(600,60)
\put(-15,48){+}
\put(-3,48){$\cdots$}
\put(15,48){+}
\put(30,54){\line(1,0){18}}
\put(30,53){\line(1,0){18}}
\put(30,52){\line(1,0){18}}
\put(30,51){\line(1,0){18}}
\put(30,50){\line(1,0){18}}
\put(30,49){\line(1,0){18}}
\put(30,48){\line(1,0){18}}
\put(30,47){\line(1,0){18}}
\put(30,46){\line(1,0){18}}
\put(60,50){\circle{25}}
\put(49,46){\line(1,1){15}}
\put(52,41){\line(1,1){17}}
\put(57.5,38.5){\line(1,1){14}}
\put(74,64){\circle{16}}
\put(79,44){\circle{16}}
\put(107.5,59){\circle{16}}
\put(64,30){\circle{16}}
\put(46,35){\circle{16}}
\put(60,62){\vector(1,3){9}}
\put(74,84){9}
\put(82,64){\vector(1,1){14}}
\put(90,80){7}
\put(82,64){\vector(3,-1){17}}
\put(92,63){1}
\put(87,46){\vector(1,1){12}}
\put(95,46){2}
\put(115.5,59){\vector(3,2){16}}
\put(125,74){1}
\put(115.5,59){\vector(3,-2){16}}
\put(125,38){2}
\put(87,46){\vector(3,-2){14}}
\put(94,29){8}
\put(65,22){\vector(1,-2){9}}
\put(78,2){3}
\put(65,22){\vector(-1,-2){9}}
\put(49,2){4}
\put(40,30){\vector(-1,-4){5}}
\put(39,8){5}
\put(40,30){\vector(-3,-2){17}}
\put(20,24){6}
\put(69,61){\small $1^{dd}$}
\put(73,41){\small $0^{dc}$}
\put(101.5,56){\small $0^{dc}$}
\put(58,27){\small $1^{uu}$}
\put(40,32){\small $1^{uu}$}
\put(59,89){$d$}
\put(102,82){$d$}
\put(100,68){$d$}
\put(86,52){$c$}
\put(137,74){$d$}
\put(137,41){$c$}
\put(105,29){$d$}
\put(79,-8){$u$}
\put(49,-8){$u$}
\put(33,-1){$u$}
\put(14,11){$u$}
\put(20,-25){$6\, I_{13}(0^{dc}1^{uu}1^{uu}1^{dd}0^{dc})\, \alpha_2^{1^{dd}0^{dc}}$}
\put(156,48){+}
\put(175,54){\line(1,0){18}}
\put(175,53){\line(1,0){18}}
\put(175,52){\line(1,0){18}}
\put(175,51){\line(1,0){18}}
\put(175,50){\line(1,0){18}}
\put(175,49){\line(1,0){18}}
\put(175,48){\line(1,0){18}}
\put(175,47){\line(1,0){18}}
\put(175,46){\line(1,0){18}}
\put(205,50){\circle{25}}
\put(194,46){\line(1,1){15}}
\put(197,41){\line(1,1){17}}
\put(202.5,38.5){\line(1,1){14}}
\put(214,68){\circle{16}}
\put(214,32){\circle{16}}
\put(253,60){\circle{16}}
\put(253,40){\circle{16}}
\put(193,34){\circle{16}}
\put(200,61){\vector(1,3){10}}
\put(213,86){9}
\put(222,70){\vector(3,2){21}}
\put(233,86){7}
\put(222,70){\vector(3,-1){23}}
\put(238,68){1}
\put(217,53){\vector(3,1){28}}
\put(238,51){2}
\put(217,47){\vector(3,-1){28}}
\put(238,42){4}
\put(222,30){\vector(3,1){23}}
\put(238,25){3}
\put(261.5,60){\vector(3,2){21}}
\put(289,71){1}
\put(261.5,60){\vector(3,-1){23}}
\put(290,52){2}
\put(261.5,40){\vector(3,1){23}}
\put(290,42){3}
\put(261.5,40){\vector(3,-2){21}}
\put(289,19){4}
\put(222,30){\vector(3,-2){21}}
\put(233,7){8}
\put(190,26.5){\vector(1,-4){5.2}}
\put(198,3){5}
\put(190,26.5){\vector(-1,-1){15}}
\put(167,9){6}
\put(208,65){\small $0^{ud}$}
\put(208,29){\small $0^{ud}$}
\put(247,57){\small $1^{uu}$}
\put(247,37){\small $1^{uu}$}
\put(187,31){\small $0^{dc}$}
\put(201,93){$d$}
\put(246,89){$d$}
\put(227,60){$u$}
\put(224,49){$u$}
\put(231,44){$u$}
\put(226,35){$u$}
\put(294,82){$u$}
\put(298,62){$u$}
\put(298,36){$u$}
\put(294,10){$u$}
\put(246,8){$d$}
\put(186,-1){$d$}
\put(172,0){$c$}
\put(170,-25){$24\, I_{14}(1^{uu}1^{uu}0^{dc}0^{ud}0^{ud})\, \alpha_2^{0^{ud}0^{ud}}$}
\put(306,48){+}
\put(325,54){\line(1,0){18}}
\put(325,53){\line(1,0){18}}
\put(325,52){\line(1,0){18}}
\put(325,51){\line(1,0){18}}
\put(325,50){\line(1,0){18}}
\put(325,49){\line(1,0){18}}
\put(325,48){\line(1,0){18}}
\put(325,47){\line(1,0){18}}
\put(325,46){\line(1,0){18}}
\put(355,50){\circle{25}}
\put(344,46){\line(1,1){15}}
\put(347,41){\line(1,1){17}}
\put(352.5,38.5){\line(1,1){14}}
\put(374,57){\circle{16}}
\put(371,37){\circle{16}}
\put(404,77){\circle{16}}
\put(401,50){\circle{16}}
\put(348,31){\circle{16}}
\put(350,61){\vector(1,4){7.5}}
\put(360,89){8}
\put(353,62){\vector(1,2){14}}
\put(372,87){9}
\put(356,62.5){\vector(3,1){40}}
\put(385,80){1}
\put(381,61){\vector(1,1){15}}
\put(392,63){2}
\put(381,61){\vector(1,-1){12}}
\put(382,46){3}
\put(379,35){\vector(1,1){14}}
\put(383,32){4}
\put(412,78){\vector(3,2){18}}
\put(435,87){1}
\put(412,78){\vector(3,-2){18}}
\put(435,64){2}
\put(409,50){\vector(3,2){18}}
\put(431,54){3}
\put(409,50){\vector(3,-2){18}}
\put(431,32){4}
\put(379,35){\vector(1,-1){16}}
\put(398,11){7}
\put(344,24){\vector(1,-4){5.5}}
\put(354,0){5}
\put(344,24){\vector(-1,-1){16}}
\put(320,7){6}
\put(368,54){\small $0^{uc}$}
\put(365,34){\small $0^{ud}$}
\put(398,74){\small $0^{dc}$}
\put(395,47){\small $1^{uu}$}
\put(342,28){\small $1^{uu}$}
\put(350,97){$d$}
\put(370,99){$d$}
\put(441,95){$d$}
\put(444,69){$c$}
\put(441,55){$u$}
\put(440,28){$u$}
\put(375,75){$d$}
\put(379,65){$c$}
\put(388,57){$u$}
\put(389,38){$u$}
\put(386,7){$d$}
\put(340,-6){$u$}
\put(322,-4){$u$}
\put(320,-25){$12\, I_{15}(0^{dc}1^{uu}1^{uu}0^{uc}0^{ud})\, \alpha_2^{0^{ud}0^{uc}}$}
\end{picture}

\vskip110pt

\begin{picture}(600,60)
\put(62,48){+}
\put(74,48){$\cdots$}
\put(92,48){+}
\put(108,54){\line(1,0){18}}
\put(108,53){\line(1,0){18}}
\put(108,52){\line(1,0){18}}
\put(108,51){\line(1,0){18}}
\put(108,50){\line(1,0){18}}
\put(108,49){\line(1,0){18}}
\put(108,48){\line(1,0){18}}
\put(108,47){\line(1,0){18}}
\put(108,46){\line(1,0){18}}
\put(138,50){\circle{25}}
\put(127,46){\line(1,1){15}}
\put(130,41){\line(1,1){17}}
\put(135.5,38.5){\line(1,1){14}}
\put(147,68){\circle{16}}
\put(158,50){\circle{16}}
\put(147,32){\circle{16}}
\put(186,60){\circle{16}}
\put(186,40){\circle{16}}
\put(125,34){\circle{16}}
\put(133,61){\vector(1,3){10}}
\put(146,86){9}
\put(155,70){\vector(3,2){21}}
\put(166,86){7}
\put(155,70){\vector(3,-1){23}}
\put(171,68){1}
\put(166,50){\vector(1,1){12}}
\put(165,56){2}
\put(166,50){\vector(1,-1){12}}
\put(165,37){3}
\put(155,30){\vector(3,1){23}}
\put(171,25){4}
\put(195,60){\vector(3,2){21}}
\put(222,71){1}
\put(195,60){\vector(3,-1){23}}
\put(223,52){2}
\put(195,40){\vector(3,1){23}}
\put(223,42){3}
\put(195,40){\vector(3,-2){21}}
\put(222,19){4}
\put(155,30){\vector(3,-2){21}}
\put(166,7){8}
\put(123,26){\vector(1,-4){5.5}}
\put(133,2){5}
\put(123,26){\vector(-1,-1){16}}
\put(99,9){6}
\put(141,65){\small $0^{ud}$}
\put(152,47){\small $0^{ud}$}
\put(141,29){\small $0^{dc}$}
\put(180,57){\small $1^{uu}$}
\put(180,37){\small $0^{dc}$}
\put(119,31){\small $1^{uu}$}
\put(133,89){$d$}
\put(177,89){$d$}
\put(164,69){$u$}
\put(173,51){$u$}
\put(174,44){$d$}
\put(163,27){$c$}
\put(229,78){$u$}
\put(231,54){$u$}
\put(231,41){$d$}
\put(229,13){$c$}
\put(177,6){$d$}
\put(120,-4){$u$}
\put(107,0){$u$}
\put(80,-40){$24\, I_{16}(1^{uu}0^{dc}1^{uu}0^{ud}0^{ud}0^{dc})\, \alpha_3^{0^{ud}0^{ud}0^{dc}}$}
\put(252,48){+}
\put(275,54){\line(1,0){18}}
\put(275,53){\line(1,0){18}}
\put(275,52){\line(1,0){18}}
\put(275,51){\line(1,0){18}}
\put(275,50){\line(1,0){18}}
\put(275,49){\line(1,0){18}}
\put(275,48){\line(1,0){18}}
\put(275,47){\line(1,0){18}}
\put(275,46){\line(1,0){18}}
\put(305,50){\circle{25}}
\put(294,46){\line(1,1){15}}
\put(297,41){\line(1,1){17}}
\put(302.5,38.5){\line(1,1){14}}
\put(299,69){\circle{16}}
\put(318,66){\circle{16}}
\put(351,62){\circle{16}}
\put(320,37){\circle{16}}
\put(303,30){\circle{16}}
\put(324,5){\circle{16}}
\put(296,76.5){\vector(-1,1){16}}
\put(275,82){5}
\put(296,76.5){\vector(1,4){5.5}}
\put(305,95){6}
\put(324,71){\vector(1,1){16}}
\put(331,87){7}
\put(324,71){\vector(2,-1){18.5}}
\put(339,68){1}
\put(317,53){\vector(3,1){25.5}}
\put(331,50){2}
\put(359,62){\vector(2,1){18}}
\put(372,75){1}
\put(359,62){\vector(2,-1){18}}
\put(372,43){2}
\put(325,30){\vector(2,-1){16}}
\put(340,27){8}
\put(325,30){\vector(-1,-4){4.2}}
\put(326,18){3}
\put(306,22.5){\vector(3,-2){14.5}}
\put(308,11){4}
\put(327,-2.5){\vector(3,-2){15}}
\put(343,-9){3}
\put(327,-2.5){\vector(-1,-3){5.7}}
\put(313,-18){4}
\put(306,22.5){\vector(-1,-3){5.7}}
\put(294,8){9}
\put(293,66){\small $1^{uu}$}
\put(312,63){\small $0^{dc}$}
\put(345,59){\small $0^{dc}$}
\put(314,34){\small $0^{ud}$}
\put(297,27){\small $0^{ud}$}
\put(318,2){\small $1^{uu}$}
\put(271,95){$u$}
\put(292,97){$u$}
\put(346,89){$d$}
\put(332,71){$c$}
\put(323,48){$d$}
\put(381,76){$c$}
\put(381,46){$d$}
\put(347,16){$d$}
\put(318,22){$u$}
\put(311,21){$u$}
\put(349,-17){$u$}
\put(327,-23){$u$}
\put(295,-4){$d$}
\put(280,-40){$12\, I_{18}(0^{dc}1^{uu}1^{uu}0^{dc}0^{ud}0^{ud})\, \alpha_3^{0^{ud}0^{ud}0^{dc}}$}
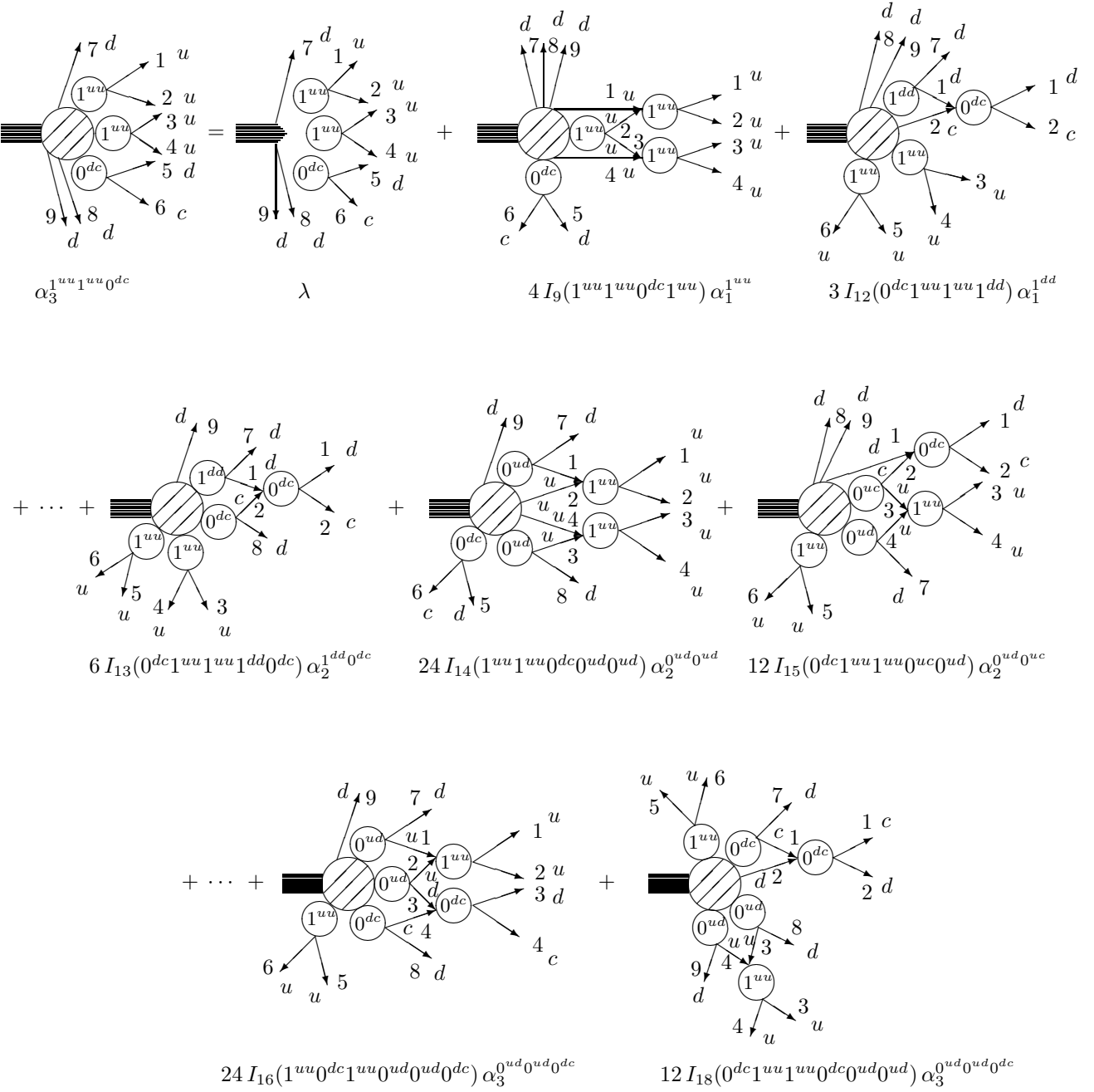
\put(0,-85){Fig. 1. The graphical equations of the reduced amplitude $\alpha_3^{1^{uu}1^{uu}0^{dc}}$
for the $ ^3_{\Sigma_c^0} He$ $pp\Sigma_c^0$ $I_3=0$ $J^P=\frac{1}{2}^+$.}
\end{picture}

\end{document}